\documentclass[prb,preprint,amsmath,amssymb]{revtex4}

\usepackage[utf8x]{inputenc}
\usepackage{graphicx}
\usepackage{float}
\usepackage{color}
\usepackage{dcolumn}
\usepackage{bm}
\usepackage{amssymb}
\usepackage{amsmath}
\usepackage{bbm}
\usepackage{epsfig}
\usepackage{psfrag}

\usepackage{etoolbox}
\makeatletter
\patchcmd{\NAT@citesuper}%
   {\textsuperscript{#1}}{\textsuperscript{[#1]}}{}{}
\makeatother

\usepackage{lineno,hyperref}

\begin{document}

\title{Enhanced Raman and photoluminescence response in monolayer MoS$_2$ due to laser healing of defects}

\author{Achintya Bera,  D V S Muthu and A K Sood\footnote{electronic
mail:asood@physics.iisc.ernet.in}}

\affiliation{Department of Physics, Indian Institute of Science, Bangalore 560 012, India}

\date{\today}

\begin{abstract}

Bound quasiparticles, negatively charged trions and neutral excitons, are associated with the direct optical transitions at the K-points of the Brillouin zone for monolayer MoS$_2$. The change in the carrier concentration, surrounding dielectric constant and defect concentration can modulate the photoluminescence and Raman spectra. Here we show that exposing the monolayer MoS$_2$ in air to a modest laser intensity for a brief period of time enhances simultaneously the photoluminescence (PL) intensity associated with both trions and excitons, together with $\sim$ 3 to 5 times increase of the Raman intensity of first and second order modes. The simultaneous increase of PL from trions and excitons cannot be understood based only on known-scenario of depletion of electron concentration in MoS$_2$ by adsorption of O$_2$ and H$_2$O molecules. This is explained by laser induced healing of defect states resulting in reduction of non-radiative Auger processes. This laser healing is corroborated by an observed increase of intensity of both the first order and second order 2LA(M) Raman modes by a factor of $\sim$ 3 to 5. The A$_{1g}$ mode hardens by $\sim$ 1.4 cm$^{-1}$ whereas the E$^1_{2g}$ mode softens by $\sim$ 1 cm$^{-1}$. The second order 2LA(M) Raman mode at $\sim$ 440 cm$^{-1}$ shows an increase in wavenumber by $\sim$ 8 cm$^{-1}$ with laser exposure. These changes are a combined effect of change in electron concentrations and oxygen-induced lattice displacements.

\vspace{15pt}

\end{abstract}

\maketitle


\section{introduction}

Monolayer molybdenum disulfide (MoS$_2$), a non-centrosymmetric semiconductor, has drawn considerable recent interest because of its technological applications, such as solar cell \cite{SCch9_2013}, photon detection \cite{PDch9_2012}, light emission \cite{LEch9_2013}, energy harvesting \cite{EHch9_2014} and spin-control optoelectronic devices \cite{SVch9_2012}. Charge impurities present in the semiconductor-metal interface for monolayer MoS$_2$ devices play an important role in strong pining of the Fermi level \cite{dass2013}. The surface defects and the defect-induced variations in local stoichiometry is responsible for the observed n-type characteristic features, irrespective of the high work function \cite{McDonnell2014}. Even the disorder or defects present in MoS$_2$ channels \cite{Ghatak2011} is a kind of intrinsic feature leading to the observed trap-assisted space charge limited current \cite{Ghatak2013}, resulting in reduced mobilities with decreasing temperature \cite{kis}.

Monolayer transition metal dichalcogenides are extremely important in developing two dimensional nanoelectronics and optoelectronics because of the presence of excitons and trions, even at room temperature \cite{makT3}. It has been proposed that the excitonic emission in monolayer MoS$_2$ can be considered in potential applications of future excitonic devices \cite{Baldo2009,Andreakou2014}. However, these bound quasiparticles (excitons and trions) are very responsive to the changes of external factors such as electrical and chemical doping, temperature, strain, surrounding medium and laser irradiation (topic of this work). The very sensitivity of MoS$_2$ to the surrounding light has been used in heterostructure \cite{Zhang2014,Roy2013} (graphene/MoS$_2$) to exhibit novel properties such as achieving photosensitivity as high as $\sim$ 10$^7$ A/W, separating electron-hole pairs at the interface. The contribution to the PL spectrum for monolayer MoS$_2$ comes from two neutral excitons (A$^0$ and B) and a negatively charged trion (A$^{-}$). The B-exciton arises due to the spin-orbit splitted valence band ($\sim$ 150 meV) \cite {XiaoPRL2012} at K-point of Brillouin zone. The PL-spectral response sensitively depends on the surrounding dielectric medium which alters the screening of the Coulomb interaction and hence the dynamics of bound quasiparticles \cite{Lin2014}. Monolayer MoS$_2$ is very prone to react with atmosphere. For example, it has been shown that by purging different gases, there is a huge enhancement ($\sim$ 100 times) in the PL spectrum as compared to the vacuum condition due to the depletion of electrons by the physisorbed O$_2$ and H$_2$O molecules \cite{Tongay2014}. It is the formation of negatively charged trions which tunes the PL-spectrum. The modulation of the PL-spectrum has been performed using electrostatic gating \cite{makT3}, by functionalizing the monolayer with cesium carbonate (Cs$_2$CO$_3$) for n-type doping to enhance the trion formation \cite{LinCesium2014}, also functionalization by p-type aromatic molecules \cite{Su2015} to get enhanced PL. The different types of substrates used to deposit exfoliated monolayer MoS$_2$, also tune the PL-spectrum \cite{Li2014}.

In optoelectronic applications, devices are exposed to light and hence it is important to know the effect of laser light on the monolayer MoS$_2$. Here, we report PL and Raman studies using 514.5 nm laser excitation (of power density $\sim$ 4.5 $\times$ 10$^5$ W/cm$^2$) on exfoliated monolayer MoS$_2$ flakes as a function of laser irradiation time (upto 25 minutes) by keeping the excitation laser in on-state. While analyzing our data, we came across a similar report by Lee et al. \cite{Lee2016} on the PL and Raman studies of monolayer MoS$_2$ (CVD-grown as compared to exfoliated monolayers in our case) as a function of laser (532 nm) exposure (of power density 5.8 $\times$ 10$^5$ W/cm$^2$) time, where three different regions were marked according to the laser irradiation time: (I) slow photo-oxidation at the initial stage (0-40 minutes) such that physisorption of ambient O$_2$ and H$_2$O molecules takes place resulting in an enhancement of PL intensity; (II) fast-oxidation at a later stage (40-50 minutes) where PL intensity increases abruptly; and (III) photoquenching (50-200 minutes) where reduction of PL intensity occurs. Because of the less laser exposure time (25 minutes), we did not observe photoquenching. Contrary to Lee at al. \cite{Lee2016}, we do observe simultaneous enhancement of trion intensity ($\sim$ 1.3 times) along with neutral excitons. Our observation on enhancement of total PL intensity (trion + A-exciton) is by $\sim$ 5.5 times, whereas in case of Lee et al. \cite{Lee2016}, it is 2 fold. According to Lee et al. \cite{Lee2016}, the Raman intensity and wavenumber increase for the A$_{1g}$ mode whereas, the E$^1_{2g}$ mode shows no change in wavenumber and intensity upto 5-minutes of laser irradiation. With more laser irradiation time, there is no further blueshift of the A$_{1g}$ mode and the Raman intensities of the A$_{1g}$ and E$^1_{2g}$ modes decrease after 5 minutes of irradiation \cite{Lee2016}. The decrease in Raman intensity was attributed \cite{Lee2016} to the degradation of the sample due to the prolonged laser irradiation. To the best of our knowledge, there is no quantitative reported data on Raman study and the dynamics of enhanced trion and B-exciton PL emission as a function of laser irradiation time. Here, we present a systematic study of time evolution (due to laser exposure) of changes in PL-spectra including B-exciton and Raman-spectra including a second-order M1-mode (attributed to 2LA(M) \cite{Kang2014,LA_Chow_2015}) at $\sim$ 440 cm$^{-1}$.  The A$_{1g}$ mode blueshifts by $\sim$ 1.4 cm$^{-1}$ and the E$^1_{2g}$ mode redshifts by $\sim$ 1 cm$^{-1}$ in the range of 0-25 minutes. The intensities of the A$_{1g}$ and E$^1_{2g}$ modes increase by $\sim$ 3 times after 25 minutes of laser irradiation. The M1-mode hardens by 8 cm$^{-1}$ and its intensity increases by $\sim$ 5 times as a function of laser irradiation time. We rule out the possibility of the degradation of the sample upto 25 minutes of laser irradiation as implied by the increment of the Raman intensity for all the observed modes. We explain the enhanced PL from trion as well as excitons based on the laser-induced annealing of defects. Lu et al. \cite{Sowch092015} reported atomic healing of defects for CVD-grown WSe$_2$ by controlled oxygen substitution of chalcogenides using focused laser (532 nm) beam of much higher power density $\sim$ 2 $\times$ 10$^7$ W/cm$^2$, as compared to our present study on exfoliated MoS$_2$ monolayer.

\section{Experimental Details}

Single-layer MoS$_2$ flakes from a bulk single crystal (SPI Supplies) were mechanically exfoliated on a 300-nm SiO$_2$/Si substrate using scotch tape. Raman and PL spectra were recorded using 514.5 nm laser excitation with a Witec confocal spectrometer. The experiments were performed at room temperature using a 50x long working distance objective. The power of laser excitation (514.5 nm) was $\sim$ 5 mW of power density corresponding to 4.5 $\times$ 10$^5$ W/cm$^2$. Raman and PL measurements were performed by keeping the laser (with constant power) in on-state and recording the time with stop-watch. The measurements times for the PL and Raman spectra were 10 and 30 seconds, respectively.

\section{Results and Discussion}

\subsection{Photoluminescence}

\begin{figure}[h!]
	\includegraphics[trim=0 0 15 70, scale=0.5]{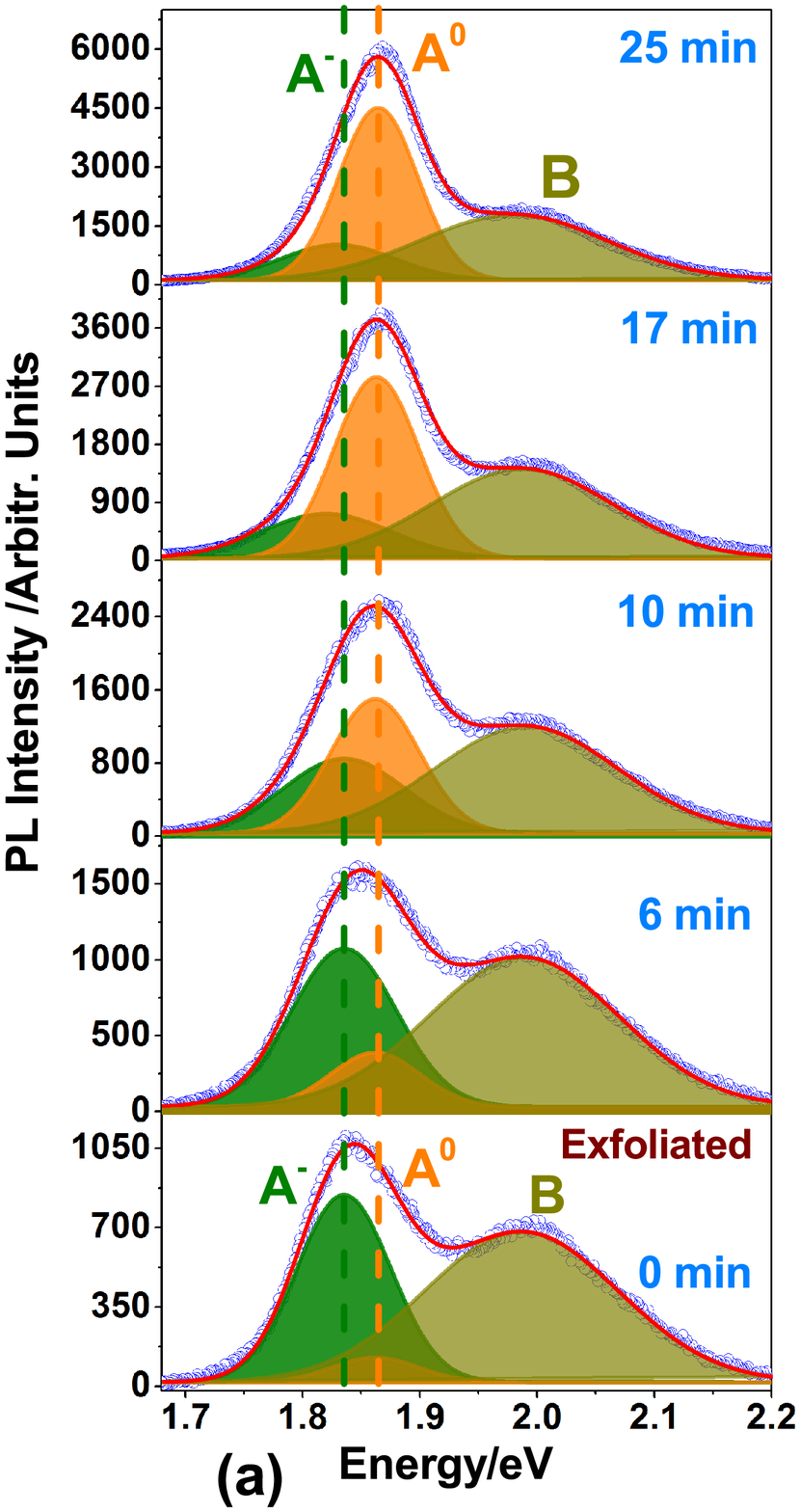}
	\includegraphics[trim=120 0 15 80, scale=0.6]{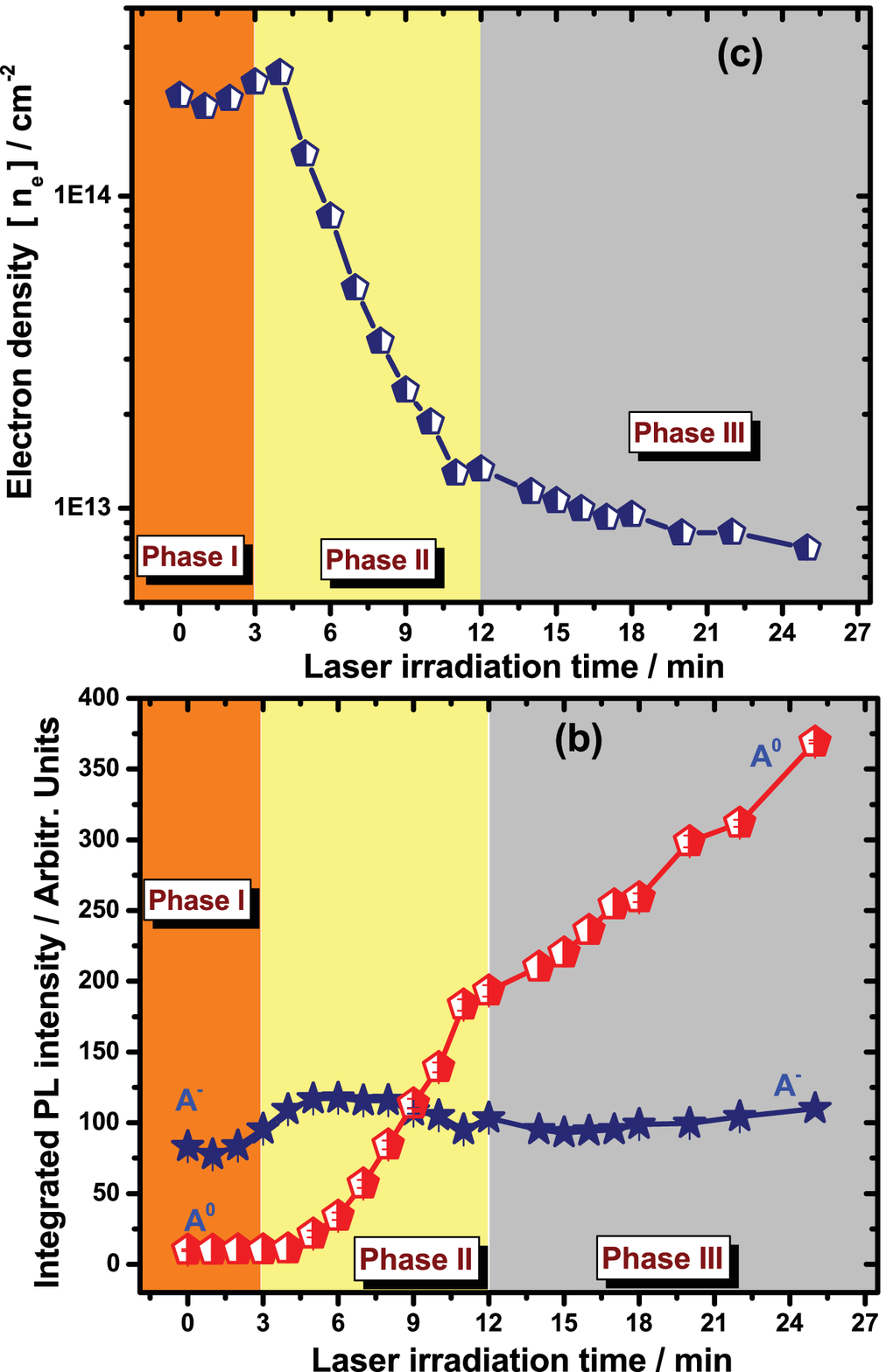}
	\caption{(a) Stacking of PL spectrum at different laser-on time in minutes (min) using 514.5-nm laser excitation. PL spectrum is fitted with three Gaussian peaks for trion (\textbf{A$^-$}), \textbf{A}-exciton  and \textbf{B}-exciton. Vertical dashed lines are guide to the eye. (b) The individual integrated intensities for trion (A$^{-}$) and for exciton (A$^0$) are plotted as a function of laser-on time. (c) The calculated electron density using Eqn~\ref{ch-09_01} is plotted. Three different shading colors mark three different regimes as discussed in text.}
	\label{ch-09_Mos2_01}
\end{figure}

Fig.~\ref{ch-09_Mos2_01}(a) shows the time evolution of PL spectra for the monolayer MoS$_2$ flake. The spectral band around 1.85 eV has been deconvoluted into negatively charged trion (A$^{-}$) and neutral exciton (A$^0$). In our analysis, the PL peak positions for A$^{-}$ and A$^0$ do not change (marked by vertical dashed lines in Fig.~\ref{ch-09_Mos2_01}a) in agreement with Lee et al. \cite{Lee2016}. The spectrum to begin with (at 0-min) is dominated by the trion-peak (A$^{-}$) at $\sim$ 1.840 eV and a very weak band associated with A-exciton (A$^0$) peaked at $\sim$ 1.865 eV.  With more exposure to the laser irradiation, the spectra are dominated by contribution from A$^0$-exciton with decreasing contribution by trion. Around 25-min, there is almost complete domination of A$^0$-exciton as indicated by vertical dashed lines.

The individual integrated intensities for trion (I$_{A^{-}}$) and for exciton (I$_{A^0}$) are plotted in Fig.~\ref{ch-09_Mos2_01}(b). The following observations can be made: (I) at the initial stage within 0-3 minutes, there is  no change in the integrated area (IA); (II) in the intermediate range of 3-12 minutes, the IA for A$^0$ exciton increases sharply, whereas A$^{-}$-trion shows a small increment. There is a cross-over between the intensities of A$^0$ and A$^{-}$ at $\sim$ 9 minutes; (III) at a later stage between 12-25 minutes, the IA for A$^{-}$ remains almost constant whereas it increases for the A$^0$ excitons. The integrated intensity of A$^{-}$ increases by $\sim$ 1.3 times in the range of 0 to 25 minutes. The integrated intensity of A$^0$ exciton is enhanced by a factor of $\sim$ 32 compared to as-exfoliated sample of MoS$_2$ and the total intensity of A$^0$ + A$^{-}$ bands is enhanced by $\sim$ 5.5 times as shown in Fig.~\ref{ch-09_Mos2_01}(a). Our observation is similar to 34 times enhancement of A$^0$ exciton by functionalizing MoS$_2$ with p-type aromatic molecules \cite{Su2015} as well as with physisorption of H$_2$O$_2$ molecules \cite{SU2015h2o2}.


The bound quasiparticles' population is a function of carrier density, i.e., n$_{A^0}$/n$_{A^{-}}$=f\{n$_{e(h)}$\}  and also is proportional to their spectral weight \cite{ylindi}. From the mass-action model, the integrated intensity carries the information of the carrier density (in our case, it is electrons) in the monolayer MoS$_2$. Mouri et al. \cite{Mouri2013} proposed a relation

\begin{equation}\label{ch-09_01}
    n_e (cm^{-2})=2.5 \times 10^{13} \times \left\{ \frac{I_{A^{-}}}{I_{A^0}}\right\}
\end{equation}

Taking measured values of integrated intensities associated with A$^{-}$ and A$^0$, the electron density is calculated as a function of laser irradiation time shown in Fig.~\ref{ch-09_Mos2_01}(c). One can observe three clear distinct regimes: (I) from 0 to 3 minutes where $n_e$ does not change significantly; (II) intermediate phase (3-12 minutes) where depletion rate of electrons is highest, followed by (III)  a slower rate of depletion of electron concentrations in the range of 12 to 25 minutes. Upon the exposure of laser irradiation, initially within 0-3 minutes the air molecules mainly O$_2$ and H$_2$O (p-types) are adsorbed on the monolayer MoS$_2$ surface \cite{Lee2016,Tongay2014} and there is no change in $n_e$. From 3 to 12 minutes, the physically absorbed O$_2$ and H$_2$O molecules start to deplete excess electrons from the monolayer \cite{Lee2016}. At a later stage where almost complete depletion of electrons happen, the PL-spectrum is dominated by neutral A-exciton (see Figs.~\ref{ch-09_Mos2_01}a and b). At this point, it is worth mentioning that it takes nearly 24 hours to come back to the trion dominated PL spectrum (same as at 0-min) after switching off the laser irradiation. By functionalization with p-type aromatic molecules, Su et al. \cite{Su2015} also reported the depletion of electrons by almost one order of magnitude (1.01 $\times$ 10$^{14}$ cm$^{-2}$ to 0.97 $\times$ 10$^{13}$ cm$^{-2}$) using Eqn~\ref{ch-09_01} and thereby enhancing the PL intensity.

\begin{figure}[h!]
	\centering
	\includegraphics[trim=80 25 55 30, scale=0.6]{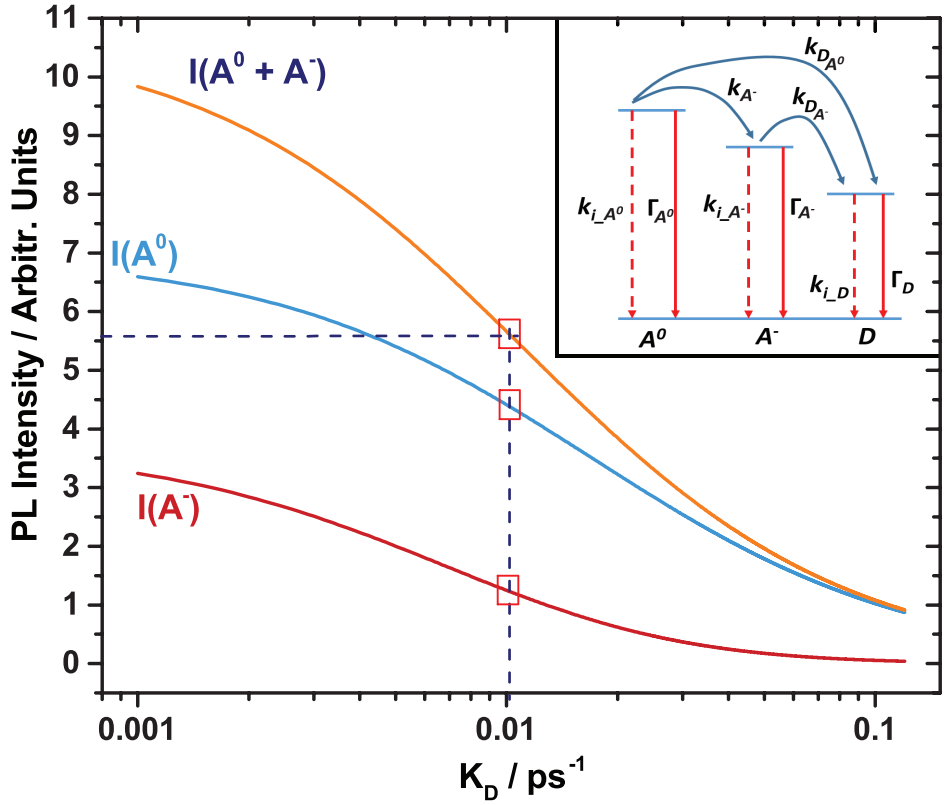}
	\caption{Calculated PL intensity vs. k$_D$ plot in semi-logarithmic scale using Eqns ~\ref{ch-09_02} and \ref{ch-09_03}. Here, I(A$^0$ + A$^{-}$)= I(A$^0$ ) + I(A$^{-}$). The dashed lines are guide to the eye. The inset \cite{Su2015} shows the schematic diagram for different radiative (solid vertical lines) and non-radiative (dashed vertical lines) recombination processes of neutral A$^0$-exciton.}
	\label{ch-09_Mos2_03a}
\end{figure}

 The time duration of different phases pertains to the kinetics of adsorption and chemisorption of oxygen molecules \cite{Nan2014} and is difficult to quantify. For example, in case of CVD-grown monolayer MoS$_2$, the phase-I for similar laser power lasts for $\sim$ 35 minutes \cite{Lee2016}, as compared to 3 minutes for the exfoliated monolayer MoS$_2$.

To understand the enhanced PL-intensity of trion along with the increased intensity of exciton, one needs to go beyond the general approach of charge transfer \cite{Tongay2014} and dielectric screening \cite{Lin2014} methods by involving defect states (D) which influence dynamics of excitons and trions. Following the model proposed by Su et al.\cite{Su2015}, the observed intensities of A$^0$-exciton and A$^{-}$-trion can be expressed in terms of their quantum yields (Q$_{A^0}$, Q$_{A^{-}}$) as,

\begin{equation}\label{ch-09_02}
    \begin{split}
        I_{A^0} & = Q_{A^0} G , \,\, and \\
        I_{A^{-}} & = Q_{A^{-}} G
        \end{split}
\end{equation}

where G is generation rate and Q$_{A^0}$ and  Q$_{A^{-}}$ are given by \cite{Su2015},

\begin{equation}\label{ch-09_03}
    \begin{split}
        Q_{A^0} & =\frac{\Gamma_{A^{0}}}{\Gamma_{A^{0}} + k_{A^{-}} + k_{D_{A^{0}}} + k_{i-A^{0}}}=\frac{0.0012}{0.0112 + k_{A^{-}} + k_D}, \,\, and \\
        Q_{A^{-}} & = \frac{k_{A^{-}}\Gamma_{A^{-}}}{(\Gamma_{A^{0}}+ k_{A^{-}} + k_{D_{A^{0}}} + k_{i-A^{0}})(\Gamma_{A^{-}} + k_{D_{A^{-}}} + k_{i-A^{-}})}=\frac{0.0012*k_{A^{-}}}{(0.0112 + k_{A^{-}} + k_D)(0.0112 + k_D)}
    \end{split}
\end{equation}

The different terms corresponding to different processes are shown in the inset of Fig.~\ref{ch-09_Mos2_03a}. Here, $\Gamma_{A^{0}}$, $\Gamma_{A^{-}}$ and $\Gamma_{D}$ represent radiative decay rates of exciton, trion and defect-trapped excitons, respectively. k$_{A^{-}}$ is decay rate of A$^0$ to A$^{-}$. k$_{D_{A^{0}}}$ and k$_{D_{A^{-}}}$ are trapping rates of A$^0$ and A$^{-}$ to D, respectively. k$_{i-A^{0}}$ , k$_{i-A^{-}}$ and k$_{i-D}$ represent non-radiative decay rates of A$^0$, A$^{-}$ and defect-trapped excitons associated with Auger process and carrier-phonon scattering \cite{Su2015}. For simplicity, $\Gamma_{A^{0}}$ = $\Gamma_{A^{-}}$ = 0.0012 ps$^{-1}$, k$_{i-A^{0}}$ = k$_{i-A^{-}}$ = 0.01 ps$^{-1}$ and k$_{D_{A^{0}}}$ = k$_{D_{A^{-}}}$ = k$_D$ are assumed \cite{Su2015}.

Fig.~\ref{ch-09_Mos2_03a} shows the calculated PL intensity for A$^0$-exciton (I$_{A^0}$) and A$^{-}$-trion (I$_{A^{-}}$) as a function of k$_D$ using Eqns ~\ref{ch-09_02} and \ref{ch-09_03}. Here, k$_{A^{-}}$ is taken to be 0.006 ps$^{-1}$ in agreement with the observed ratio of I$_{A^0}$/I$_{A^{-}}$=3.0 (see Fig.~\ref{ch-09_Mos2_01}b) at 25 minutes. The simultaneous enhancements of A$^{-}$-trion intensity by $\sim$ 1.3 times along with the increased intensity of A$^0$-exciton are marked by red-boxes corresponding to k$_D$ $\approx$ 0.01 ps$^{-1}$ as shown in Fig.~\ref{ch-09_Mos2_03a}. The enhancement of calculated total intensity I(A$^0$ + A$^{-}$) by $\sim$ 5.5 times (see Fig.~\ref{ch-09_Mos2_03a}) corroborates the observed enhancements of A$^0$ + A$^{-}$ bands (see Fig.~\ref{ch-09_Mos2_01}a). Fig.~\ref{ch-09_Mos2_03a} qualitatively supports the fact that the simultaneous enhancement of trion and A$^0$-exciton is possible when the rate constant k$_D$ related to defect states reduces. We note that this observation is quite distinct from the studies by Tongay et al. \cite{Tongay2014}, Lin et al. \cite{Lin2014}, Mouri et al. \cite{Mouri2013} and Lee et al. \cite{Lee2016} where, the enhanced PL-intensity is associated with the enhancement of A$^0$-exciton along with the decreased trion intensity. Our observation of enhanced trion-intensity is similar to the studies by Su et al. showing simultaneous enhancements of trion-intensity ($\sim$ 3.5 times) along with A$^0$-exciton for the melamine doped \cite{Su2015} monolayer MoS$_2$ as well as functionalizing with H$_2$O$_2$ molecules \cite{SU2015h2o2}, and it was attributed to the defects-assisted dynamics of bound quasiparticles. In our case, we propose that defect states created by S-vacancies can be effectively healed with laser irradiation in air through the adsorption of O$_2$ and H$_2$O molecules which will decrease the rate constant k$_D$. As mentioned earlier, here k$_D$ represents the rate of trapping of excitons and trions to the defect states \cite{Su2015}. The sulphur vacancies can be passivated in presence of laser irradiation by adsorption of oxygen (O$_{ad}$) as explained by Lu et al. \cite{Sowch092015}. Here, the capture of oxygen by S vacancy is an exoenergetic processes, with an enthalpy balance of -4.7 eV. This is larger than the binding energy per atom of the oxygen molecule (- 2.5 eV) and hence the oxygen atom gets substituted in place of S vacancy, without introducing any midgap states \cite{Sowch092015,Ataca2011}. In addition, the oxygen can be adsorbed on top of S atom, forming O$_{ad}$-S configuration. This configuration is lower in energy than the O$_{ad}$-Mo configuration.

\begin{figure}[h!]
	\centering
	\includegraphics[trim=20 0 55 0, scale=0.9]{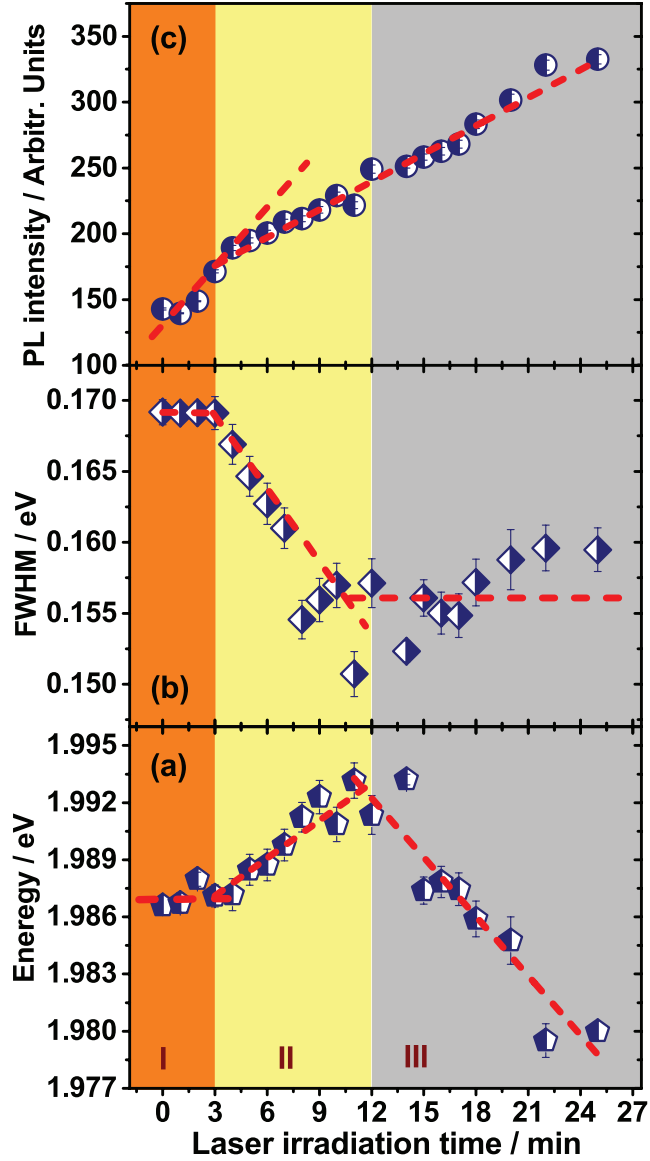}
	\caption{(a) PL peak position, (b) FWHM and (c) integrated area are plotted for B-exciton. Three different shading colors mark three different regimes as discussed in text. The dashed lines are guide to the eye. Note that the arbitrary units of integrated intensity in (c) are same as in Fig.~\ref{ch-09_Mos2_01}(b).}
	\label{ch-09_Mos2_03}
\end{figure}

We have used the model to explain the simultaneous enhancement of PL intensities associated with trions and excitons. The increase in I$_{A^0}$ and I$_{A^{-}}$ in phase-II is much faster than that in phase-III. This implies that the rate of laser-induced defect healing is faster in phase-II as compared to phase-III. This kinetics is not been included in the model. The model predicts the intensities of excitons and trions, given a certain density of defects (via the value of k$_D$).

The PL peak position, FWHM (full width at half maximum) and the integrated area for the B-exciton are plotted in Fig.~\ref{ch-09_Mos2_03}(a),(b) and (c), respectively. Initially, there is no change in the PL-peak position; in the intermediate range, it is blueshifted by $\sim$ 6 meV ; and at later stage, it is redshifted by $\sim$ 14 meV (see Fig.~\ref{ch-09_Mos2_03}a). We suggest that the shift of the peak position of exciton is a combined effect of doping and possible changes in the bandgap due to defect-healing. The depletion of electrons will enhance the Coulomb interaction due to less screening and thus will increase the binding energy of the exciton, leading to the redshift of the peak position. On the other hand, the lattice relaxation due to defect passivation can renormalize the bandgap to higher value and hence leads to the blueshift.

 \begin{figure}[h!]
	\centering
	\includegraphics[trim=50 30 45 50, scale=0.52]{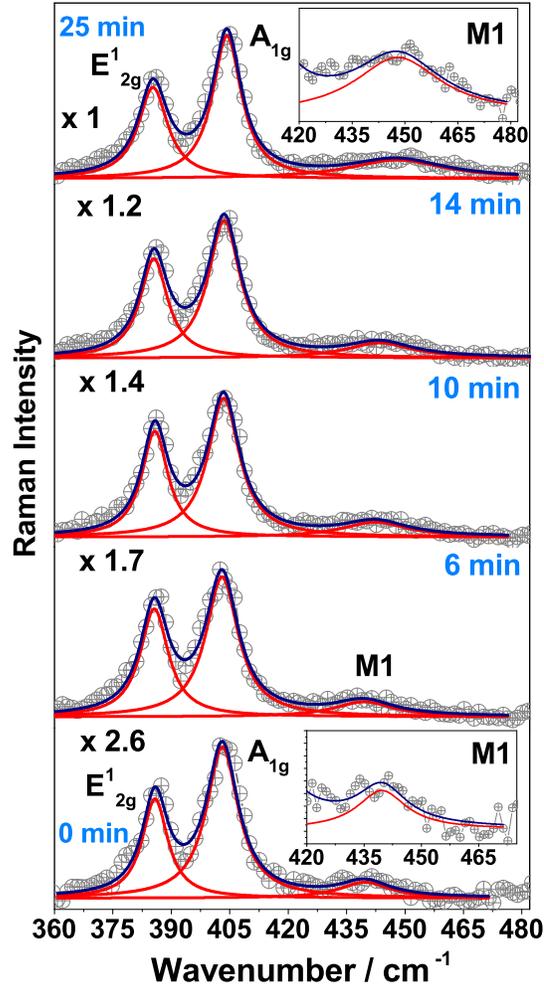}
	\caption{Evolution of Raman spectra as a function of time. Solid lines are the Lorentzian fits to the experimental data points. The insets at 0-min and 25-min show the M1 mode in zoomed-scale. The number in the corresponding spectrum represents the scale enhancement factor.}
	\label{ch-09_Mos2_04}
\end{figure}

For the FWHM (see Fig.~\ref{ch-09_Mos2_03}b), there is no change in the phase-I whereas it continually decreases in phase-II and then remains almost unchanged in phase-III. Our observation of the decrease in FWHM might be related to the reduced exciton-carrier scattering \cite{Li2014} as more depletion of electrons occur. The integrated area increases with laser irradiation as shown in Fig.~\ref{ch-09_Mos2_03}(c). Note that the arbitrary units of integrated intensity in Fig.~\ref{ch-09_Mos2_03}(c) are same as in Fig.~\ref{ch-09_Mos2_01}(b). The increase of integrated intensity of A$^0$-exciton along with the B-exciton are related to the enhanced recombination process as well as reduced Auger process \cite{Su2015} arising from laser-induced healing of defects.


 \subsection{Raman scattering as a function of laser irradiation}

\begin{figure}[h!]
	\centering
	\includegraphics[trim=50 0 45 0, scale=0.8]{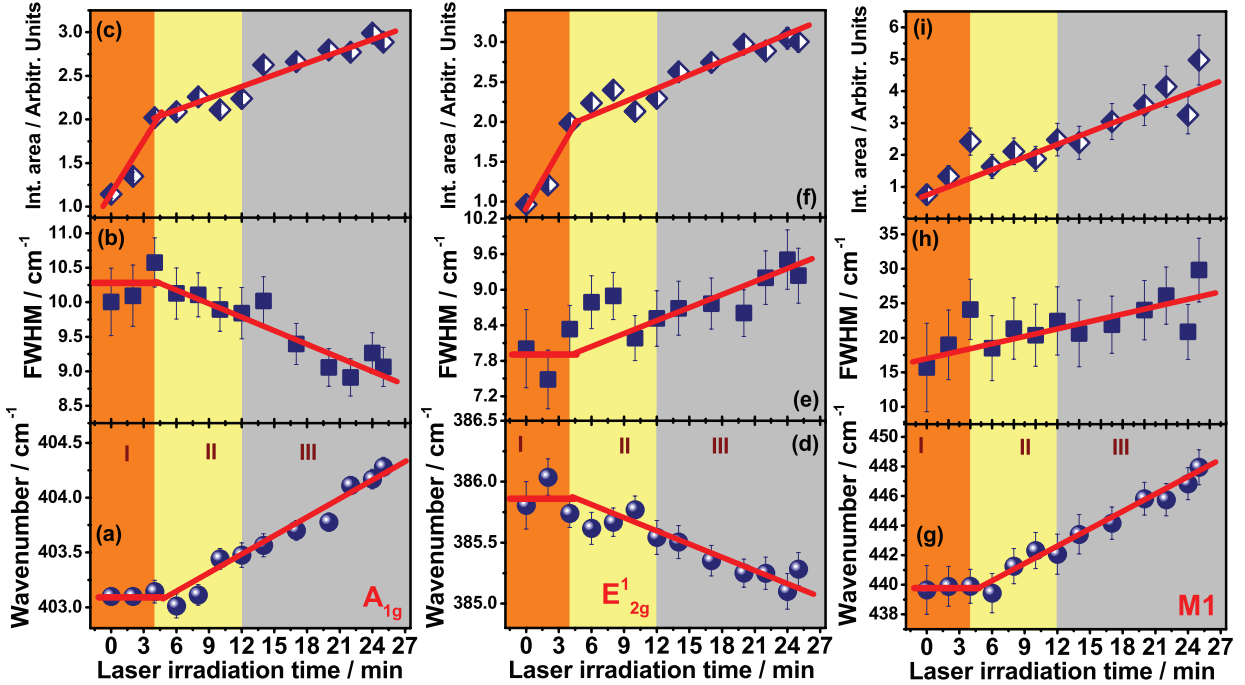}
	\caption{ 514.5 nm laser excitation is used. Wavenumber, FWHM and integrated area are plotted for the A$_{1g}$ mode in (a),(b) and (c); for the E$^1_{2g}$ mode in (d),(e) and (f); for M1 mode in (g), (h) and (i), respectively. The solid lines are guide to the eye. The error bars are also plotted and can be smaller than the size of the symbol.}
	\label{ch-09_Mos2_05}
\end{figure}

Raman spectroscopy is a very powerful tool to probe not only the layer number but also the doping concentration in monolayer MoS$_2$ \cite{berabook2014,bera}. It also gives information about structural--related distortions or damages of the sample on the length scale of laser spot size. Fig.~\ref{ch-09_Mos2_04} shows the time evolution of Raman spectra with 514.5 nm laser excitation on the same monolayer MoS$_2$ flake. Raman spectra are recorded using the same power density (4.5 $\times$ 10$^5$ W/cm$^2$) with total accumulation time of 30 seconds. The M1 mode ($\sim$ 440 cm$^{-1}$) was assigned \cite{Kang2014,LA_Chow_2015} as a second order longitudinal acoustic mode (2LA) at the M-point of Brillouin zone. The quantitative variations of the wavenumber, FWHM and the integrated area are plotted in Fig.~\ref{ch-09_Mos2_05} as a function of laser-irradiation time for all the observed Raman modes. The wavenumbers of A$_{1g}$ and E$^1_{2g}$ modes do not change in phase-I (see Figs.~\ref{ch-09_Mos2_05}a and d), followed by hardening of the A$_{1g}$ mode by $\sim$ 1.4 cm$^{-1}$ and softening of the E$^1_{2g}$ mode by $\sim$ 1 cm$^{-1}$ in region II and III. The FWHM remains constant at initial stage and then gradually decreases for A$_{1g}$ mode by $\sim$ 1.5 cm$^{-1}$ as shown in Fig.~\ref{ch-09_Mos2_05}(b), whereas for the E$^1_{2g}$ mode, it increases (see Fig.~\ref{ch-09_Mos2_05}e). The increment in the integrated area by $\sim$ 3 times of the A$_{1g}$ and E$^1_{2g}$ modes are shown in Figs.~\ref{ch-09_Mos2_05}(c) and (f), respectively. The effect of laser irradiation on the second-order mode M1 ($\sim$ 440 cm$^{-1}$) is shown in Figs.~\ref{ch-09_Mos2_05}(g),(h) and (i) for the wavenumber, FWHM and the integrated area, respectively. In phase-I, there is no change in wavenumber and after that it hardens by $\sim$ 8 cm$^{-1}$. The M1 mode shows $\sim$ 5 times ($\sim$ 400 \%) enhancement in the integrated intensity as shown in Fig.~\ref{ch-09_Mos2_05}(i).

Now we address how to understand the observed changes in Raman modes as a function of laser irradiation. The E$^1_{2g}$ mode involves anti-parallel vibration of two S atoms with respect to the Mo atom in the basal plane while the A$_{1g}$ mode has out-of plane vibration of only S atoms in opposite directions \cite{berabook2014}. Chakraborty et al. \cite{bera} showed that electron doping decreases the A$_{1g}$ mode wavenumber without affecting the E$^1_{2g}$ mode. This can explain the increase of the A$_{1g}$ phonon wavenumber with laser irradiation (i.e. decreasing n$_e$, see Fig.~\ref{ch-09_Mos2_01}c). The decrease in wavenumber of E$^1_{2g}$ mode with laser irradiation shown in Fig.~\ref{ch-09_Mos2_05}(d) cannot be explained by reduced n$_e$. We note that oxygen plasma treatment \cite{Kang2014,Ye2016NL} of monolayer MoS$_2$ shows Raman results similar to our data (Figs.~\ref{ch-09_Mos2_05}a and d). Therefore, the blueshift of the A$_{1g}$ mode with laser irradiation time can be associated with the depletion of electrons from the monolayer MoS$_2$ and the softening of E$^1_{2g}$ mode is associated with the oxygen-induced displacements of Mo and S atoms from their original sites \cite{Kang2014}. This point needs further theoretical work. The decrease in FWHM for the A$_{1g}$ mode is associated with the depletion of electrons, since the coupling between electron and phonon can strongly influence the A$_{1g}$ phonon in monolayer MoS$_2$ and it is extremely sensitive to the electron concentration in the channel \cite{bera}. The large increase in wavenumber ($\sim$ 8 cm$^{-1}$) of the M1 mode might be related to the strong electron-phonon coupling for the corresponding zone-edge acoustic phonons (LA$_K$) at K-point of Brillouin zone \cite{LiLAch092013} and it needs further study.

The increase in integrated area for all the Raman modes by $\sim$ 3 to 5 times is associated with the reduced screening due to depletion of electrons and laser-annealing of defects. This also shows that in our experiments, the sample is not damaged till 25 minutes of irradiation as the intensity of all the Raman modes increase.

\section{Conclusions}

In summary, the monolayer MoS$_2$ is very sensitive to the laser irradiation in air such that it helps to deplete the extra electron density and thereby enhancing the photoluminescence intensity. The defect states plays a crucial role in realization of simultaneous enhancement of trion-intensity along with the increment of A$^0$-exciton intensity and hence, laser irradiation serves as a practical way in healing of defects in monolayer MoS$_2$.  This is further corroborated by $\sim$ 3--5 times increment of Raman intensity for all the Raman active modes, along with the signatures of reduced electron-doping. We hope that our experimental findings will help to understand about how laser exposure can change the optoelectronic properties in designing the devices as well as in sensor applications with materials like MoS$_2$.

\section{Acknowledgments}

AKS acknowledges the funding from Department of Science and Technology and JC Bose National Fellowship, India. AB thanks CSIR for research fellowship. \\

\end{document}